# Simultaneous Optical and Meteor Head Echo measurements using the Middle Atmosphere Alomar Radar System (MAARSY): Data collection and preliminary analysis

P. Brown*[1,2], G. Stober[3], C. Schult[3], Z. Krzeminski[1], W. Cooke[4], J.L. Chau[3]

[1]Dept. of Physics and Astronomy, University of Western Ontario, London, Ontario, Canada N6A 3K7

[2]Centre for Planetary Science and Exploration, University of Western Ontario, London, Ontario, Canada N6A 5B8

[3]Leibniz Institute of Atmospheric Physics, Rostock University
    Kühlungsborn, Germany

[4]NASA Meteoroid Environment Office, EV44,
    Marshall Space Flight Center, Huntsville, AL, USA

*Corresponding authors email: pbrown@uwo.ca

44 pages
3 tables
12 figures

**Proposed Running Head** : Simultaneous Optical-MAARSY meteor measurements

Editorial Correspondence to :

Dr. Peter Brown
Department of Physics and Astronomy
University of Western Ontario
London, ON
N6A 3K7
CANADA

Phone : 1-519-661-2111 x86458

Fax : 1-519-661-4085

E-mail address : pbrown@uwo.ca

Highlights
- 105 optical meteors simultaneously detected as head echoes by MAARSY are analysed
- Radiants measured by radar and optical show median differences of 1.5 degrees.
- Optical calibration shows MAARSY detects meteoroids of masses $10^{-9}$ kg - $10^{-10}$ kg
- Clear trend of larger RCS for brighter meteors at higher heights and larger speeds
- Many events show variations in RCS significantly different than optical light curve

## Abstract

The initial results of a two year simultaneous optical-radar meteor campaign are described. Analysis of 105 double-station optical meteors having plane of sky intersection angles greater than 5 degrees and trail lengths in excess of 2 km also detected by the Middle Atmosphere Alomar Radar System (MAARSY) as head echoes was performed. These events show a median deviation in radiants between radar and optical determinations of 1.5 degrees, with 1/3 of events having radiant agreement to less than one degree. MAARSY tends to record average speeds roughly 0.5 km/s and 1.3 km higher than optical records, in part due to the higher sensitivity of MAARSY as compared to the optical instruments. More than 98% of all head echoes are not detected with the optical system. Using this non-detection ratio and the known limiting sensitivity of the cameras, we estimate that the limiting meteoroid detection mass of MAARSY is in the $10^{-9}$ kg to $10^{-10}$ kg (astronomical limiting meteor magnitudes of +11 to +12) appropriate to speeds from 30-60 km/s. There is a clear trend of higher peak RCS for brighter meteors between 35 and -30 dBsm. For meteors with similar magnitudes, the MAARSY head echo radar cross-section is larger at higher speeds. Brighter meteors at fixed heights and similar speeds have consistently, on average, larger RCS values, in accordance with established scattering theory. However, our data show RCS $\propto$ v/2, much weaker than the normally assumed RCS $\propto$ v$^3$, a consequence of our requiring head echoes to also be detectable optically. Most events show a smooth variation of RCS with height broadly following the light production behavior. A significant minority of meteors show large variations in RCS relative to the optical light curve over common height intervals, reflecting fragmentation or possibly differential ablation. No optically detected meteor occurring in the main radar beam and at times when the radar was collecting head echo data went unrecorded by MAARSY. Thus there does not appear to be any large scale bias in MAARSY head echo detections for the (comparatively) larger optical events in our dataset, even at very low speeds.

# 1. Introduction

Determination of the fundamental characteristics of meteoroids remains a central goal of all meteor measurements. The size, mass, bulk density and chemistry of meteoroids may be inferred from measurements of their ablation behavior when coupled to appropriate models. However, direct measurements of these fundamental quantities during the brief period of meteoroid ablation is generally not possible. Direct meteor measurements include light production, electron production or shock production as a function of time and height (Ceplecha et al., 1998). Ultimately these "collisional" by-products of meteoroid ablation can be used individually or collectively to estimate the original mass, bulk density and size of the progenitor meteoroid, but always with assumptions. For example, ablation is often assumed to proceed without significant large scale fragmentation a simplification termed single-body ablation. However, meteoroid fragmentation has long been recognized as ubiquitous at all meteoroid sizes (Jacchia, 1955; Ceplecha et al., 1993), with recent estimates suggesting >90% of mm-sized meteoroids undergo fragmentation (Subasinghe et al., 2016). As the details of the fragmentation process remain poorly constrained (Campbell-Brown et al., 2013), validation of the differing technique-specific assumptions in deriving mass from deceleration (dynamic mass), optical brightness (photometric mass), shock production (infrasonic mass) or electron production (ionization mass) can best be achieved by simultaneous multi-technique measurements of the same meteor. Moreover, entry models may use multi-technique measurements to better constrain forward models of meteoroid ablation (eg. Silber et al., 2015).

Here we describe a two-year campaign of simultaneous automated meteor optical and head echo radar measurements conducted with the Middle Atmosphere Alomar Radar System (MAARSY). The goals of this campaign include:

1. Compare speeds, begin/end heights and radiants of head echoes measured by MAARSY and two station optical solutions for a range of meteoroid masses.

2. Compare photometric and dynamic mass measured optically with radar-derived masses with the goal of intercalibration of mass scales.

3. Use the best observed simultaneous events to fuse all metric, photometric and ionization estimates together and apply different ablation models to self-consistently model these highest quality events.

In this work we focus on results related to goal 1 to validate MAARSY trajectory and speed estimates using multi-station optical data. Additionally, statistics of the observed begin and end height of MAARSY compared to optical measurements are used to gain physical insight into the mechanisms of head echo scattering and associated biases. We briefly examine goal 2 using measured photometric light curves to compare with radar cross section (RCS). Parts of goal 2 and all of goal 3 are addressed in the companion paper by Stober et al (2017).

## 1.1 Techniques and Challenges in Meteoroid Mass Measurements

Dynamic mass refers to the instantaneous mass found when measuring meteoroid deceleration and applying the standard single body meteoroid drag equation:

$$M_d = \frac{\Gamma A}{\left(\frac{dV}{dt}\right)} \rho_a V^2 \tag{1}$$

where $V$ is the speed of the meteoroid, $\Gamma$ is the drag coefficient (usually assumed to be unity in free molecular flow conditions – eg. Campbell-Brown and Koschney, 2003), $A$ is the cross-sectional area of the meteoroid, $M_d$ is the meteoroid mass estimated via dynamics and $\rho_a$ is the atmospheric mass density and $\rho_m$ the meteoroid bulk density. Deceleration is difficult to measure, particularly with optical techniques. However, head echo observations using high power large aperture (HPLA) radars have better time and spatial (range) resolution than optical systems and may directly measure Doppler speeds, allowing routine measurement of decelerations even for fast meteors (eg. Evans, 1966; Chau and Woodman, 2004; Li and Close, 2016). The fundamental assumption when applying this equation to meteor measurements to estimate mass is that the object is non-fragmenting and ablates in a self-similar manner while keeping all other parameters ($A$, $\rho_m$) constant. The non-fragmentation assumption is a poor one (Babadzhanov, 2002). As a result, direct dynamic mass estimates tend to be systematically too small relative to photometric mass estimates, an effect long recognized in optical observations of larger meteoroids and ascribed to fragmentation (Ceplecha and ReVelle, 2005).

Photometric mass ($M_p$) estimates are not as susceptible to mis-interpretation due to fragmentation. Light production (I) depends on the rate of mass ablation and not the character of the ablation such that:

$$M_p = \frac{2}{\tau V^2} \int I dt \qquad (2)$$

where $\tau$ is the fraction of kinetic energy converted to light in the sensor passband. Here we ignore that small contribution due to the deceleration of the meteoroid to light production, a generally valid assumption for small, higher speed meteoroids (Ceplecha et al., 1998). The limiting factor in obtaining accurate $M_p$ measurements is that $\tau$ is poorly known and likely depends on speed, height, flow regime, mass and meteoroid chemistry (Jones and Halliday, 2001; Campbell-Brown et al., 2013) in addition to being pass-band specific.

A mass estimate for larger (>cm-sized) meteoroids is possible if an infrasonic signal from the meteor is detected at the ground (eg. Edwards, 2010). In this case, ignoring deceleration and assuming a cylindrical shock wave, the infrasonic mass loss ($dm_I$) over the trail length $dL$ of the meteoroid generating the observed ground signal is given by:

$$dm_I = 2\left(\frac{R_o}{V}\right)^2 P_o dL = \frac{2E_o dL}{V^2} \quad (3)$$

where $R_o$ is the cylindrical blast radius, defined as $(E_o/P_o)^{0.5}$ where $E_o$ is the meteoroid energy deposition per unit trail length and $P_o$ is the atmospheric pressure at the source altitude (Silber et al., 2015). $R_o$ is difficult to measure, though it may be estimated from the infrasonic signal amplitude or period using analytic weak-shock models (Edwards, 2010) or more accurately from computational fluid dynamic modelling (Henneton et al., 2015; Aftosmis et al 2016). In practice, equation 3 only provides an estimate of the mass lost in the ~kilometer of trail nearest the specular point of the meteor as seen from a single ground location. An integral form of eq (3) could be constructed to give the total initial meteoroid mass using a dense array of infrasound sensors on the ground to reconstruct $R_o$ vs height, but this is impractical.

Finally, meteoroid mass can be computed based on the production of electrons during ablation in the form:

$$dm_q = -\frac{q\mu V}{\beta} dt \quad (4)$$

Here $dm_q$ is the instantaneous mass computed based on measurement of the number of electrons per unit trail length ($q$) and $\mu$ is the average atomic mass of the ablated meteoroid atoms. The coefficient $β$ is the ionization probability or equivalently the number of electrons produced per ablated atom of a specific meteoric species. Theoretical and experimental estimates for $β$ made in recent years (Jones, 1997; Thomas et al., 2016) show reasonable agreement, suggesting that while uncertainties remain in the value for $β$ (particularly at low speeds) it is arguably better understood and constrained than is $τ$. The electron line density can be found at the specular point from transverse backscatter meteor observations (Weryk and Brown, 2013) by measuring received echo power. However, this can only be done if the initial meteor trail radius and radial electron distribution are known (or assumed) by applying full-wave scattering solution to find the reflection coefficient (see Poulter and Baggaley (1977) for a complete description). Unfortunately, the initial trail radius remains poorly known and measurements of the initial radius show large scatter (Jones and Campbell-Brown, 2005). This limits the accuracy of $q$ measurements using this technique, which further suffers from the limitation that $q$ is only measured at one point of the trail, so only ($dm_q$) at the specular point is determined. As with infrasonic mass, if an array of receiver stations were deployed on the ground to sample many specular points along the trail a complete profile of $q$ vs. height might be constructed. Such ionization curves from transverse scattering measurements have been crudely constructed in the past (Verniani, 1966) but high fidelity measurements are difficult.

Using radio waves radially scattered from the meteor head, a complete ionization profile may be constructed (Baggaley, 2002). Head echo observations yield the total initial meteoroid mass ($M_q$) via:

$$M_q = \int \frac{q \mu V}{\beta} dt \qquad (5)$$

In general, head echo observations produce a received power from the point (head echo) target which can be converted, using the hard target radar equation (Kero et al., 2008) to radar cross section and ultimately to electron line density using an appropriate scattering model (eg. Close et al., 2005; Marshall and Close, 2015). The radial distribution of electrons about the meteor head and the electron cloud radius must be known for these models. This is a source of uncertainty in this method, but multi-frequency observations can provide some constraints (Close et al., 2004; Marshall et al., 2016). The scattered power from each head echo depends on the interplay between the number of electrons produced from ablation (which decreases at higher altitudes but increases with mass and speed) and the size of the electron cloud which scales with the atmospheric mean free path. As shown in Close et al (2007) the resulting altitude vs. RCS profiles tend to show a maximum around 100-110 km (though this is frequency dependent), which is where head echoes are most abundant (Janches et al., 2014; Schult et al., 2016).

## 1.1 Previous Optical-radar Simultaneous meteor Studies

Various past experimental studies have attempted simultaneous optical-radar meteor observations; a recent summary is given in Weryk and Brown (2013). Here we describe only those studies which performed simultaneous head echo – optical measurements.

Nishimura et al (2001) compared two nights of simultaneous electro-optical single station meteor measurements with head echoes detected by the MU radar. They found a log-linear relation between the peak absolute optical brightness of 34 meteors with the associated head echo power, albeit with significant scatter. They showed that at least some of their meteor light curves had similar shapes to the radar head echo power curves. However, they had to presume both peaked at the same time as their optical measurements did not have independent absolute time calibration.

Michell (2010) analysed seven single-station optical meteors detected as head echoes by the Poker Flat Incoherent Scatter Radar (PFISR). He showed that the optically brighter meteors tended to be correlated with larger received echo power and that many of the bright meteors were detected in radar side-lobes.

Campbell-Brown et al (2012) deployed two station intensified optical systems to perform common-volume observations with the EISCAT radar. Four meteors with good geometry were detected by both optical stations and EISCAT during an 11 hour campaign. The optical speeds agreed with the radar determined speeds within uncertainty bounds. The absolute trajectories deviated by slightly more than respective errors, but were similar and differed by absolute values of order a hundred meters. Photometric masses and head echo ionization-derived masses were found to agree within a factor of several, despite large uncertainties in luminous and ionization efficiencies.

Most recently, Michell et al (2015) described in detail six optical counterparts to head echoes recorded by the Southern Argentine Agile Meteor Radar (SAAMER). Among this dataset four events had single station optical speeds within 10 km/s of head echo estimates, while origin directions showed a larger relative spread. The photometric mass was found to correlate with head echo SNR. Absolute masses found to agree between the two techniques in most cases to better than a factor of two.

## 2. Campaign, Hardware and Data collection

Two automated optical stations were established in September, 2014 to coordinate observations with MAARSY near the Andøya rocket range in Norway. The campaign builds on the experience of a comparable optical-radar campaign using the EISCAT radar (Campbell-Brown et al., 2012) where it became clear that to obtain significant head echo – optical meteor statistics a long duration optical and radar simultaneous measurements were needed.

To meet these criteria, MAARSY is ideal as an HPLA head echo system as it began semi-continuous fully automated head echo observations in late 2013 (Schult et al., 2016). The two automated optical stations are located at the ALOMAR observatory (69.2783°N, 16.0093°E) and 15 km south at Saura (69.1410°N, 16.0169°E ). The ALOMAR station is located ~2 km from the MAARSY site and the optical cameras are pointed to overlap the main vertically directed beam of MAARSY at an altitude of 100 km. Because of the short optical baseline, significant overlap with the main MAARSY beam and the two optical cameras occurred everywhere above 80 km altitude. Figure 1 shows the relative spacing of the optical stations and MAARSY.

Optical observations were made when darkness, cloud and lunar conditions allowed. No data were collected when the moon was above the horizon, regardless of lunar phase. Automated detection of cloud and aurora at each site was made continuously and the intensified camera (see later) not engaged unless all conditions were favorable.

During the campaign two modes of optical data recording occurred in parallel. First, a triggered system was employed whereby all optical data from both sites were cutout from large video data buffers within 5 seconds centered around the time of automatically detected head echoes cued from MAARSY. Only the 20 strongest (by SNR) head echoes each hour were used for these triggered cutouts. Secondly, the wide field cameras used the ASGARD detection software (Weryk et al., 2008) to detect any bright (magnitude brighter than +4) meteor events. This optical detection trigger resulted in video data from all cameras being saved for later comparison to radar data.

Due to the high northerly latitude of Andøya (+70° N), optical observations are only possible between September and April. In 2014, measurements began on Oct 3, 2014 and the final night of data collection was Apr 14, 2015. In all, 73 hours of useable optical data were recorded at the ALOMAR site and 116 hours at Saura during the 2014 campaign year. In total 108 optical events were simultaneously detected at both stations and as head echoes by MAARSY. In 2015, the earliest data collection occurred on Sep 15, 2015 and the last was on Apr 8, 2016. In total 242 hours of useable optical data were collected at Alomar and 369 h from Saura. A total of 140 multi-station meteors were detected optically at both stations and showed clear association with MAARSY head echoes.

## 2.1 Radar Hardware, Data Collection and Reduction Procedures

MAARSY is a high-power-large aperture radar with an active phased array antenna on the Norwegian Island of Andøya (69.3°N, 16°E). MAARSY uses a multi-channel receiver system allowing the selection of 16 different antenna sub-arrays from amongst the full 433 antennas for interferometric purposes. The hardware and technical details of MAARSY are summarized in Table 1. A more detailed description of the radar itself can be found in Latteck et al. (2012).

The earliest meteor head echo measurements with MAARSY were conducted during the ECOMA sounding rocket campaign in 2010 (Stober et al. 2013, Schult et al. 2013). Since this campaign the system has been upgraded several times. Most notably MAARSY now consists of circularly polarized antennas. Additional receiving channels (16 in total) were also implemented expanding the ambiguity area up to the third side lobe (~ +-16° off zenith); that is head echoes detected up to 16° from the zenith can be uniquely located within the beam without angular ambiguities.

The experimental mode for MAARSY used for the simultaneous radar-camera observations is shown in Table 2. The experimental parameters were selected as a compromise to permit observation of other mesospheric echoes. This experiment mode started in November 2013 and is used for continuous mesosphere monitoring, allowing also for nearly continuous head echo detection. However, the experiment runs in a sequence with other experiments and system control algorithm resulting in some time gaps (2-6 seconds) between successive observation records. For the joint radar-optical campaign the experiment used a 16 bit complementary transmit pulse code. To detect meteor head echoes the raw MAARSY data stream is separated into an undecoded pulse to pulse raw voltage stream which are then searched for head echoes. Simultaneously, a stream of decoded data with additional coherent integrations is produced for atmospheric measurements before final storage of all the data. It is the undecoded meteor head echo raw voltage data stream which is searched for likely events on site and only potential head echoes are kept. Due to the fast velocities of the meteoroids and, hence, the fast change of target range, the two alternating code sequences are kept in separate data streams and the decoding is also performed separately for each of the complementary codes. Therefore the code range side lobes are much larger than would be the case for standard complementary codes or the single Barker codes. The decoding of the Doppler-shifted code pairs is done similar to the technique described in Kero et al. (2012).

We determine the position of the meteor head echo within the radar beam for every radar pulse using the interferometric capabilities of the radar system. The analysis follows the general procedure given in Schult et al. (2013), but uses more receiver sub-arrays following the hardware upgrade post-2013. Once all pulses from a head echo are found, the shortest baselines (which correspond to a 15.6° interferometric ambiguity) are used for a first guess of the mean trajectory. A second step refinement to this solution uses the largest baseline to then determine a more precise head echo trajectory. The quality of the interferometric measurements is ensured by regular phase calibrations using radio sources which pass through the beam (eg. Chau et al., 2014).

From each recorded pulse from a meteor head echo we obtain the three dimensional position (from range and interferometry), the Signal-to-Noise Ratio (SNR), the radar cross section (RCS), and the Doppler (radial) velocity on a pulse to pulse basis. Combining these points together using a non-linear least squares fitting routine, we can then find the absolute velocity vector at each pulse and the radiant of the meteor. The resulting trajectory within the radar beam and the orbital parameters are then estimated in a second step. The statistical uncertainties of the estimated parameters, in particular, for the azimuth, elevation and velocity depend on the trajectory length and the corresponding SNR values and vary for every meteor head echo. All events that are recorded by the cameras as well as by MAARSY are manually quality controlled or manually reprocessed in order to ensure the best quality for all estimated parameters.

## 2.2 Optical Hardware, Data Collection and Reduction Procedures

Each optical site had identical equipment, software and operations. Figure 2 shows the camera hardware at each station. There are two cameras at each station which are co-aligned. Initially (Sep 2014 – Feb, 2015) these were situated beneath plastic auroral domes (Figure 3a), but the optical quality of these domes was not sufficient for the narrower field, higher resolution camera. As a result, flat BK7 glass windows were installed at both stations in Feb, 2015, greatly improving optical quality (Figure 3b).

The wide-field (14x11 degree), lower resolution 8-bit NTSC frame rate Watec Ultimate H2 camera was digitized using a Hauppauge Impact VCB Model #00558. This camera operated at all times when the sun was below the horizon and functions to detect cloud, aurora and LIDAR (at the ALOMAR station) as well as to record meteors to an approximate limiting magnitude of +5. Using the conditions from the wide field camera, an automated module determined if it was clear and dark such that the narrow field image intensified camera (which can be damaged if the illumination in the field is too high) could be activated. This narrow field intensified camera had a circular field of view of 6° and was imaged by an AVT GX1050 14-bit digital camera with a 1k x 1k CCD at 50 frames per second to a limiting meteor magnitude of roughly +7. Figure 4a and 4b shows a stacked video image of a meteor captured by the WATEC (wide-field) camera at the ALOMAR site and the same meteor recorded by the narrow field intensified system at the Saura site.

At each site one computer is used to control both cameras and time stamps are generated on each frame by NTP protocols with a precision better than the interframe rate. Video data are stored in a lossless format and astrometric and photometric reduction are performed manually on each event post-detection. The astrometric and photometric video analysis procedures follows the methodology described in Weryk and Brown (2012; 2013) and Weryk et al (2013). The astrometric stellar fits in the wide-field cameras show average residuals of 0.01°-0.02° while the narrow field-intensified system had average stellar residuals in the plate fits of 0.002°.

In this reduction process, the location of each meteor is manually measured on each frame at each site using a centroid-assisted method which provides frame-to-frame pick consistency for the leading edge of the meteor. A manual mask is constructed at each frame which captures all the meteor luminosity and the background is then subtracted from the sum of the pixel intensities across this mask. Calibration is done for each event using field stars in the G (Gaia – Jordi et al, 2010) passband and the uncalibrated log sum pixel intensities of the meteor per frame is then converted to an apparent (calibrated) magnitude. The pixel intensity for the Watec cameras is scaled according to the intensity dependent dual-gamma model of Ehlert et al (2016). The astrometric measurements from both sites are combined in a non-linear least squares solver to determine the best-fit path (Borovicka, 1990) and the light curve for each meteor from each station is then computed as a function of both height and time from each station. Absolute magnitudes are computed (normalized to 100 km range) and the light curve from each station compared. We find peak magnitude differences of less than 0.2 magnitude between stations, with small differences usually due to differences in the photometric calibrations. An example light curve is shown in Figure

5. To compute absolute luminosity we take the G-band zero magnitude to be 945W following Weryk and Brown (2013).

Due to the poor quality of the optical domes for the first half of the campaign, no useable two station narrow field, intensified meteor events were recorded until the fall of 2015.

## 4. Results and Discussion

The radar head echo trajectory and the optical two station trajectory are first compared by using the independent, but presumably common height estimate, from each instrument as a function of time and then computing an equivalent time shift in the video solution required to match height vs. time measured by MAARSY. We found that most two station optical detections with good intersection geometry were systematically between 1-2 frames earlier than the radar time. This systematic offset is expected as the NTSC framegrabber at the computer writes the frame time at the end of each frame acquisition, producing a minimum one frame bias not including latency from the camera to the disk. In contrast, the narrow field, digital cameras, which are timestamped at the camera at the start of the frame, had a median value of less than a frame of offset from the radar timing using common heights from both systems as a calibrator.

For each simultaneous head echo and multi-station optical event, the apparent pulse-to-pulse location of the head echo based on interferometry and height was overlain on the stacked meteor image from each site. This provided a direct visual estimate of the apparent goodness of fit between the head echo and plane of sky optical trajectory (Figure 6). In all cases, the head echo trajectory and optical trajectory are parallel, but a (random) small offset is commonly observed which may represent small phase errors from the MAARSY radar. It is noteworthy that a very similar (though slightly larger) offset was reported by Nishimura et al (2001). The number of standalone two station narrow field intensified events is small (~20) and so we do not analyze statistics for these events (where meteors typically move out of the field of view in any case, so we have only partial trails). Rather these data are used in a future paper to refine modelling fits between the optical data and MAARSY.

Of the 248 Watec events detected at both stations and showing a head echo with MAARSY, 105 showed well-behaved head echo solutions and could be fully analyzed on optical records from both sites. These included many events with poor geometry (ie. 28 had intersecting plane angles between the apparent plane of the meteor less than 5 degrees) or short trail lengths (14 had optical trail lengths less than 2 km, translating into only 4-5 frames per site). Computing the apparent radiant from all optical data (no filter quality for selection) we find a median deviation in radiants between the head echo estimate for radiant and the optical radiant of 1.5 degrees. Fully 1/3 show agreement in radiant direction to better than one degree. Filtering for bad geometry and short trail lengths removed most of the outliers, though some large deviations remained, in some cases due to low SNR head echo trajectories or short optical trail lengths.

A slight asymmetry in the mean speeds measured by radar as compared to the optical solutions is evident in the dataset as a whole. Figure 7 shows the distribution of mean speeds measured optically minus the average Doppler speed measured by MAARSY. There is a skew in the speeds such that MAARSY tends to record average speeds roughly 0.5 km/s above the optical values. This is a consequence of MAARSY typically recording common head echoes higher than the optical systems and hence before significant deceleration. This is shown in Figure 8 where the difference in first recorded height between the optical camera solutions and MAARSY initial head echo detection height is shown. On average MAARSY detects meteors 1.3 km higher than the optical cameras, though there are significant numbers of events where the optical systems detect the event first, usually when ablation starts outside the main beam. At higher altitudes and/or nearer the center of the beam, this offset becomes more pronounced demonstrating that MAARSY is more sensitive to the start of ablation than the cameras. That MAARSY has a lower limiting sensitivity than the optical cameras is also underscored by the fact that more than 98% of all head echoes detected in the MAARSY main beam are not detectable at all by the sensitive narrow field cameras, despite having interferometric locations which should place these many (fainter) head echoes in the optical field.

Using these number statistics, we may also compute an approximate minimum limiting meteor magnitude for MAARSY. As the limiting meteor magnitude for the narrow field system is approximately +6 to +7 and assuming a sporadic population index of 2.5, we get that the number of head echoes MAARSY sees relative to the camera sensitivity gives a bound to the limiting MAARSY sensitivity near magnitude +11 to +12. Using the mass – magnitude-speed relation of Verniani (1973) this corresponds to meteoroid masses in the $10^{-9}$ kg to $10^{-10}$ kg range for typical raw head echo speeds of 30-60 km/s.

Examination of the absolute optical magnitude at the height where MAARSY records the peak radar cross section for each of the 105 head echoes in our dataset shows clear correlations. Figure 9 shows the RCS versus magnitude at the peak RCS height as a function of speed and height at the peak RCS . There is a clear trend of both higher peak

RCS with brighter meteors and for meteors with similar magnitudes, the RCS is much larger for higher speeds as expected. The apparent drop in detectable speed at fainter magnitudes reflects the large variation in optical camera sensitivity with angular speed (eg. Hawkes, 2002). Figure 10 shows the height versus peak RCS as a function of speed and absolute brightness. The trend of brighter meteors at fixed heights and similar speeds having larger RCS' is even more noticeable. This trend is consistent with established scattering theory which predicts that for a fixed plasma radius and similar electron radial distribution, the RCS will scale with total number of electrons for plasma radii large compared to the radar wavelength and that this is also proportional to total light production (e.g. Marshall and Close, 2015).

However, our data suggests this is not the complete picture. Figure 11 shows four examples of the trends in RCS with magnitude as a function of height. We find that over half of all our events with good SNR and where RCS measurements are not made near the edge of the beam show light curves and RCS variations which are similar in shape, as would be expected ignoring the effects of head echo radii "compression" at lower heights. Examples of this behavior are shown in Figure 11 (top two plots). However, a significant minority of good events (where RCS are computed well away from beam edges) show large oscillations in RCS relative to the optical light curve for the same height intervals. Examples of this class of event are shown in Figure 11 (bottom two plots). These events may reflect fragmentation producing multiple scattering centers and hence interference in the received echo signal or perhaps differential ablation. The latter, however, if present, has no optical signature. More detailed modelling of specific events are needed.

To further explore the RCS variation with speed (and height – noting that these two variables are correlated) as a function of brightness, in Figure 12 we show all points for all 59 head echoes which were within 5 degrees of the centre of the MAARSY beam. This angular cutoff avoids any large corrections to the RCS which occur when the head echo gets close to the first beam null near ~7 degrees. Individual head echoes can be followed as nearly vertical lines. A general decrease in RCS at lower heights and lower speeds is evident, but significant scatter in RCS at near constant height and speeds are apparent. Using these data and binning the mean RCS values per 5 km/s bins we find that RCS $\propto$ v/2. We find a similar result using the RCS vs. speed applying the same analysis to the 105 events where we examined only the brightness at the peak RCS . This would seem to be at variance with theory and other HPLA observations using scattering models (eg. Close et al (2007)) which suggest that RCS $\propto$ $v^3$. The underlying reason for our different dependence is that we require the head echoes to also be detected by optical systems which have a strong bias toward slower meteors due to trailing losses. As a result, the velocity distribution of our simultaneous optical-radar events (Figure 13) does not have the typical high-speed peak present for head echo data (Close et al 2007).

This underscores the requirement that to fully interpret these data a detailed head echo scattering model generating RCS and/or electron line density estimates as a function of height and speed appropriate to MAARSY's operating wavelength is required. Moreover this model needs to be applied per event as our dataset is biased by the requirement that head echoes are also detectable optically. At present, the only published model estimates with this level of detail are for ALTAIR measurements at 160 MHz (Close et al., 2007).

Finally, we optically registered 510 common two station meteors with the ASGARD detection software which did not correlate with any automatically registered head echoes in the MAARSY dataset. For these events, we found that 225 occurred entirely outside the main beam and therefore were less likely to be detected and/or measured with accurate interferometry. A further 78 had most of their optical trajectory outside the main beam and occurred near the null (if inside the beam) making detection unlikely. The remaining 207 had optical trajectories that were mostly in the main beam and should have been detected by MAARSY. We examined each of these events manually. Our manual examination showed that all of these events were either visible in MAARSY data (but not detected by the automated algorithm either due to severe fragmentation confusing the head echo tracking algorithm or had low SNR) or occurred in gaps between radar acquisition cycles. We did not find a single optical two station meteor which appeared in the MAARSY main beam which was not visible as a head echo to the radar for other than explainable (i.e. between acquisition cycles) reasons. We found some of these optical events with speeds as low as 11-12 km/s still showed up as (weak) head echoes in MAARSY data.

## 5. Conclusions

We have detected 105 double-station optical meteor events simultaneously observed by MAARSY as head echoes. Metric comparisons establish that the absolute interferometry for MAARSY is in generally excellent agreement with optical meteor trajectories, though a slight systematic offset of a few tenths of a degree is commonly noted. We find a median deviation in radiants between radar and optical radiants of 1.5 degrees, with 1/3 of events having radiant agreement to less than one degree. MAARSY tends to record average speeds roughly 0.5 km/s above the optical values as the head echoes also are detected, on average, 1.3 km higher than the first optical registration. This difference is larger near the center of the beam demonstrating that MAARSY is more sensitive to the start of ablation than the cameras. The good agreement in two station optical speeds and trajectory suggests we may confidently compare single station optical detections with MAARSY in future and hence extend photometric – RCS correlations to fainter magnitudes than is possible with two station detections

Using just the ratio of the number of common optically detected to not detected head echoes we find that the MAARSY head echo limiting meteoroid mass lies in the $10^{-9}$ kg to $10^{-10}$ kg for speeds from 30-60 km/s

There is a clear general trend of higher peak RCS with brighter meteors as predicted by scattering theory. However, we find that RCS $\propto v/2$ at variance with theory which suggest that RCS $\propto v^3$ emphasizing the strong bias in our data toward slower meteors more easily detected optically. A significant number of common head echo – optical meteors show large variations in RCS relative to their optical lightcurve over common height intervals.These events may reflect fragmentation or possibly differential ablation.

No optically detected meteor occurring in the main radar beam and at times when the radar was collecting head echo data went unrecorded by MAARSY. Thus there does not appear to be any large scale bias in MAARSY head echo detections for the (comparatively) larger optical events, even at very low speeds.

## Acknowledgements

PGB thanks the Canada Research Chair program. This work is supported in part by the Natural Sciences and Engineering Research Council of Canada and the NASA Meteoroid Environment Office through co-operative agreements NN61AB76A and NN65AC94A. Helpful discussions with S. Close and R. Marshall are gratefully acknowledged. This work was supported in part by grant STO 1053/1-1 (AHEAD) of the Deutsche Forschungsgemeinschaft (DFG).

Tables.

Table 1. Hardware specifications for MAARSY.

| Hardware Specification | |
|---|---|
| Frequency | 53.5 MHz |
| Transceiver-modules | 433 |
| Power | ~866 kW |
| Antennas | 433 3-element (crossed) Yagi Antennas |
| Gain | 33.7 dBi |
| Aperture | ~6300 m$^2$ |
| Beam width (3dB points) | 3.6° |
| Beam steering capabilities | freely steerable with 35° off-zenith |
| Receiver channels | 16 |

Table 2. MAARSY radar experiment details for optical campaign data collection (Sep, 2014-April, 2016).

| **Experiment Specification** | |
|---|---|
| Pulse Repetition Freq. | 1000 Hz |
| Pulse coding | 16-bit complementary |
| Pulse length | 4.8 km (32 μs) |
| Duty Cycle | 3.2% |
| Range Resolution | 300 m |
| Start Range | 49800 m |
| End Range | 134700 m |
| Beam direction | vertical (zenith pointing) |
| | |

Table 3. Optical system specifications.

| | |
|---|---|
| Wide Field Cameras | Watec Ultimate H2 (768x508 pixels) |
| Narrow Field Cameras | AVT Prosilica GX1050 (1000x1000 pixels) |
| Bit Depth | Watec (8 bit); AVT (14 bit) |
| Field of View | Watec (14°×11°); AVT (6° diameter) |
| Frame Rate | Watec (30 fps – 2:1 interlaced); AVT (50 fps - progressive) |
| Limiting Stellar Magnitude (per frame) | Watec (+6); AVT (+10) |
| Approximate Meteor Limiting Magnitude | Watec (+5); AVT (+7) |

Figure Captions.

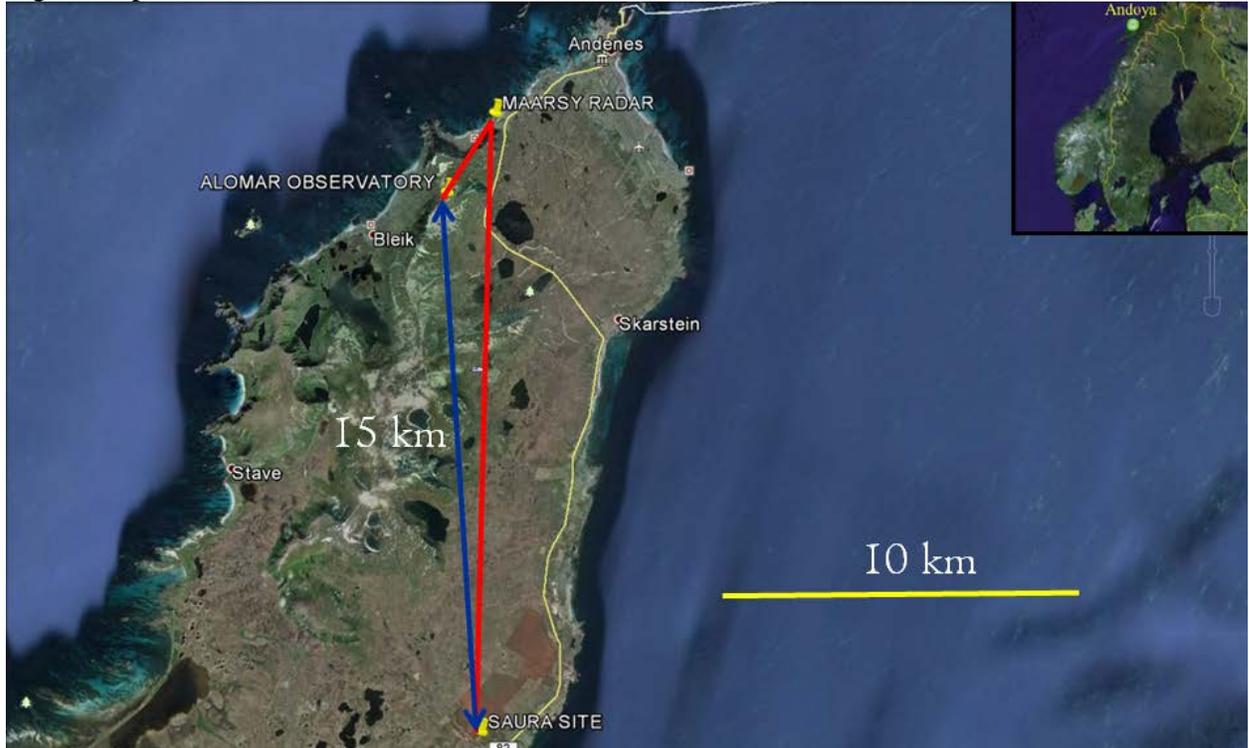

Figure 1. Relative location and baseline (blue line) of the two optical stations at the Alomar Observatory (top) and the Saura HF radar site. The location of the MAARSY radar is shown as is the pointing azimuth of the cameras (red lines). Also shown are the Scandanavian countries on the inset map for context.

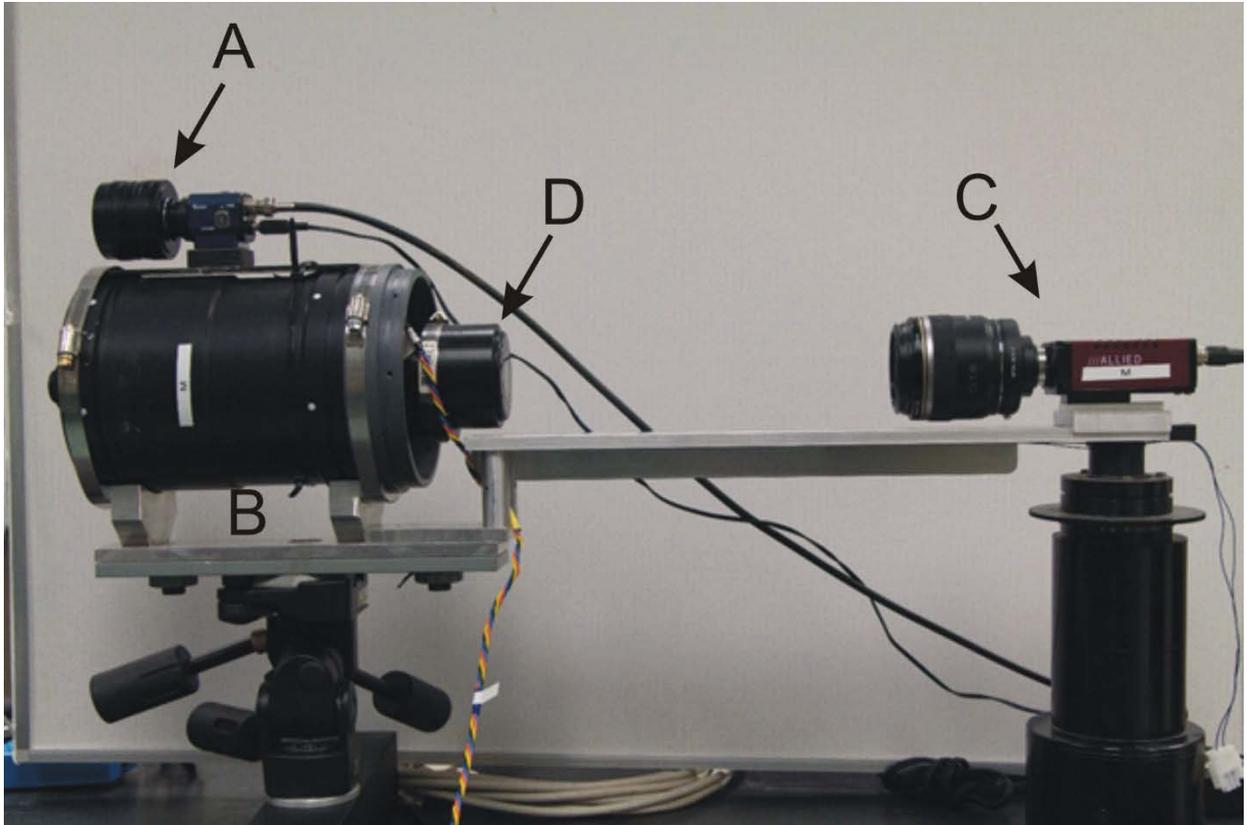

Figure 2. Optical camera system installed at each station. A – Watec Ultimate H2 camera with EXView HAD CCD using 25 mm f/0.85 fast video lens. This unit is aligned co-linearly with a 155mm f/1.2 catadioptric lens (TVS-5) (B) with $2^{nd}$ generation Litton 25 mm MCP image intensifier having an S20 photocathode (D). The intensifier output phosphor is imaged using an AVT Prosillica GX1050 camera equipped with a Canon EF-S 60mm f/2.8 Macro lens (C).

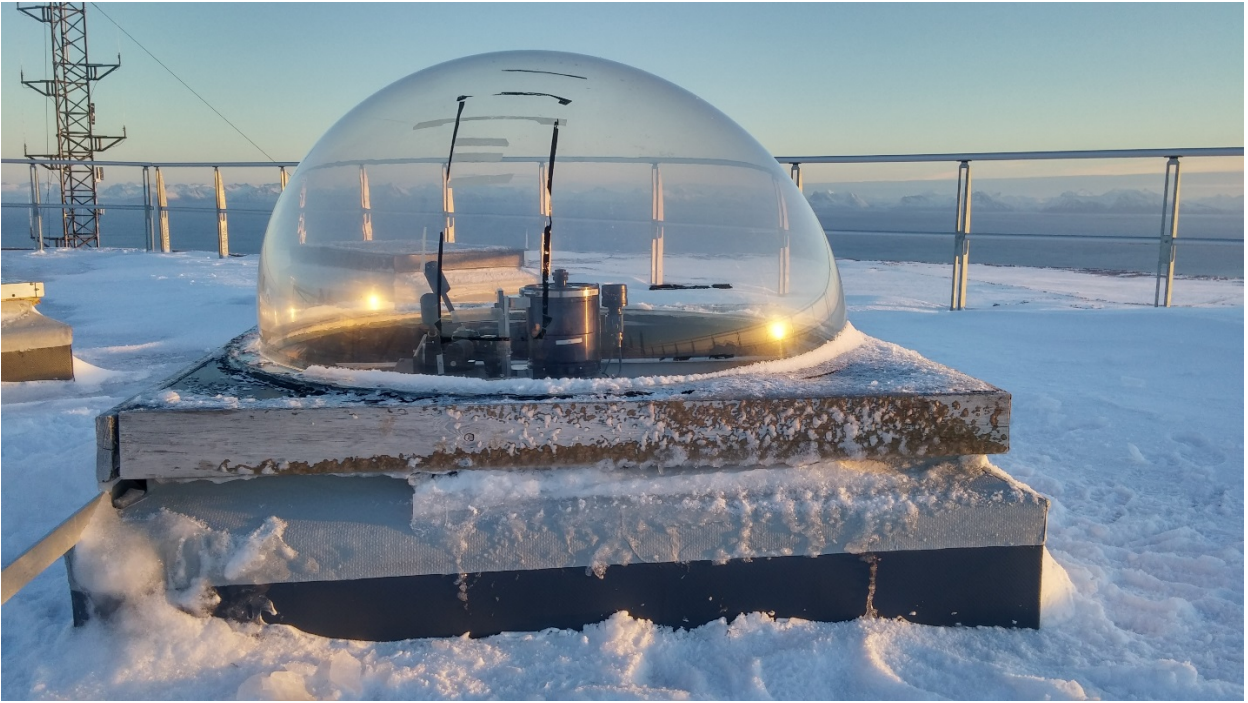

Figure 3a (top). Original installation of optical cameras under auroral dome at the Alomar site.
Figure 3b (bottom). Newly installed flat plane BK7 optical window (left of auroral camera dome) at the Saura site from the side and (inset) from above.

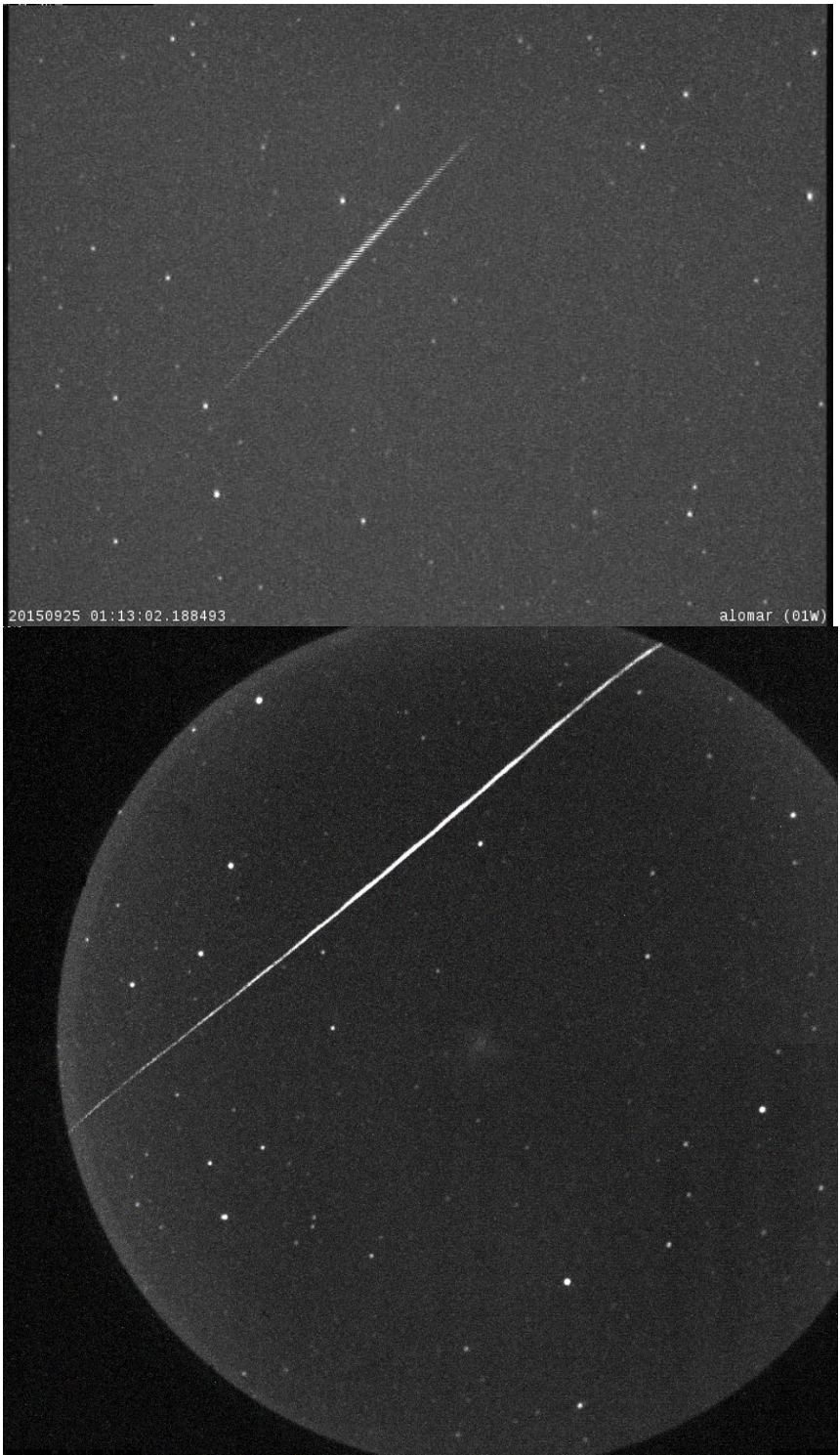

Figure 4a (top). Stacked image from the WATEC camera of a meteor of +2 peak magnitude detected at 011302 UT on September 25, 2015 from ALOMAR.
Figure 4b (bottom). The same meteor detected by the narrow field intensified camera at the Saura site.

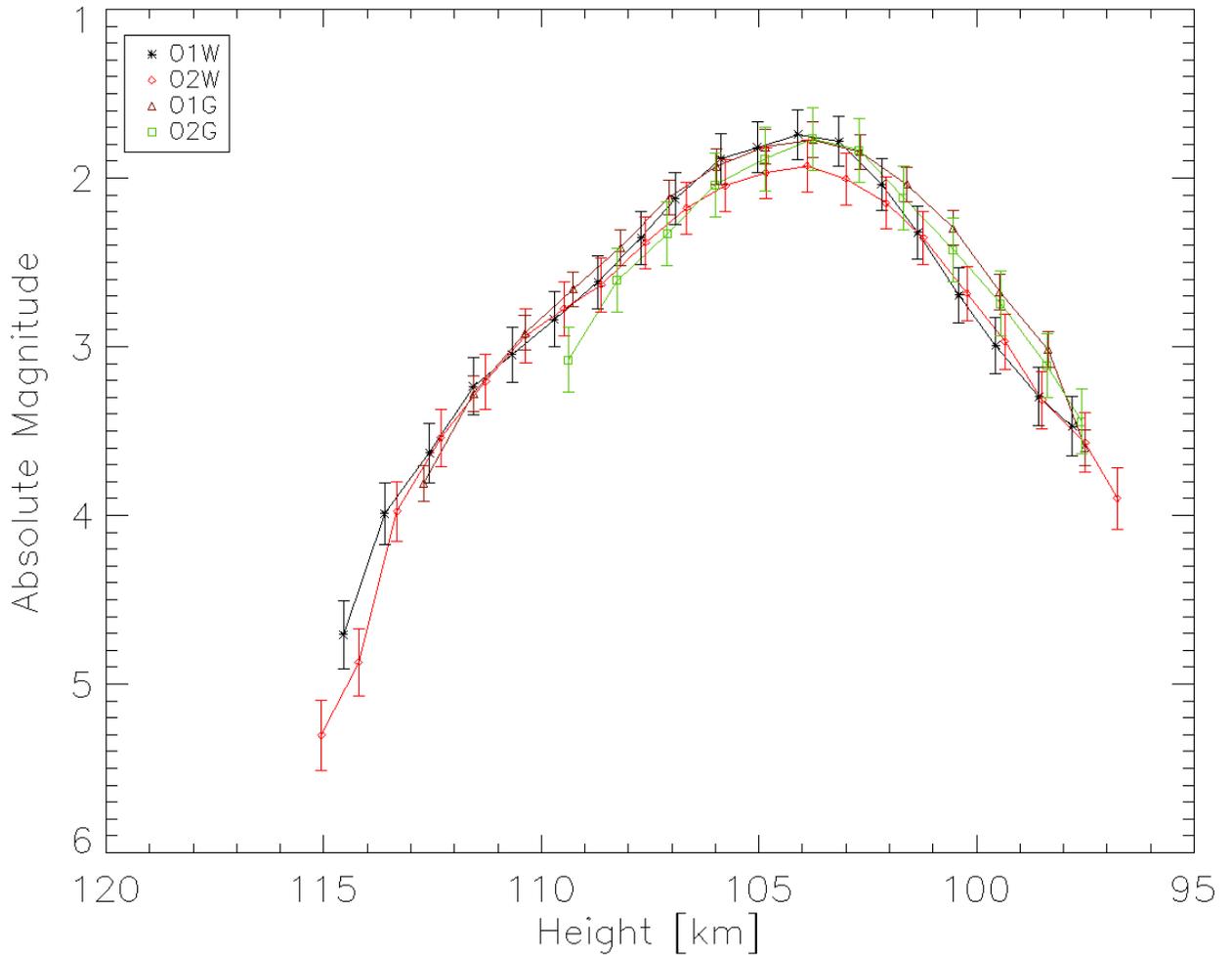

Figure 5. Calibrated G-band lightcurve for the 011302 UT meteor on September 25, 2015 as detected by the WATEC camera at Alomar (01W – black asterisk), the narrow-field intensified camera at Alomar (01G – red triangles), the Watec camera at Saura (02W – red diamonds) and the narrow-field intensified camera at Saura (02G – green squares). The error in each data point is dominated by the uncertainty in the photometric calibration, except at the faintest magnitudes where photon statistics become important.

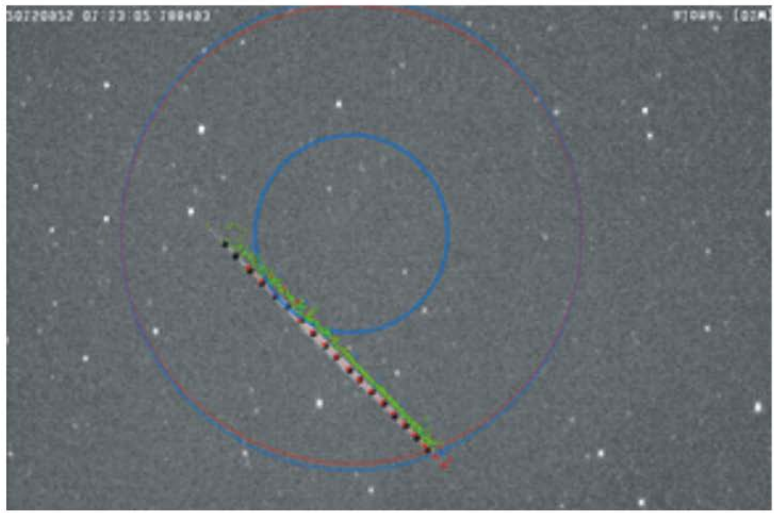
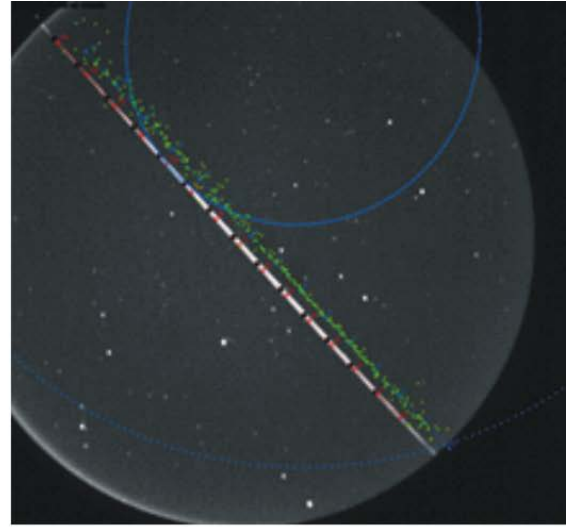
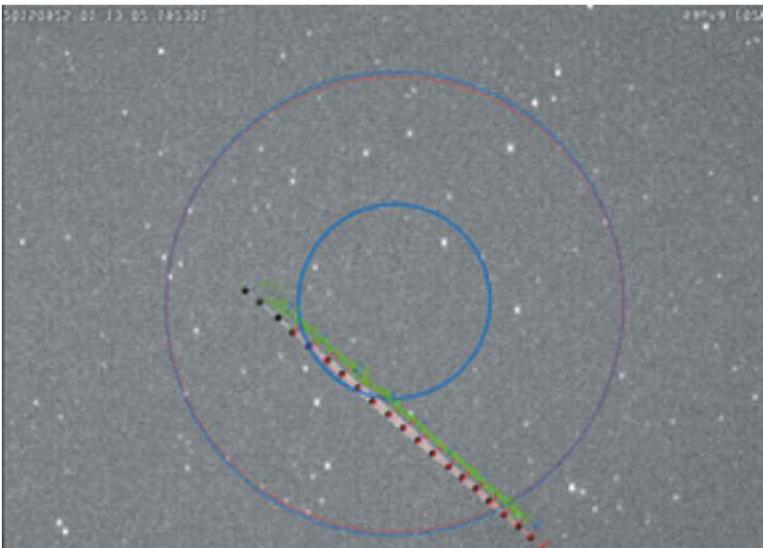
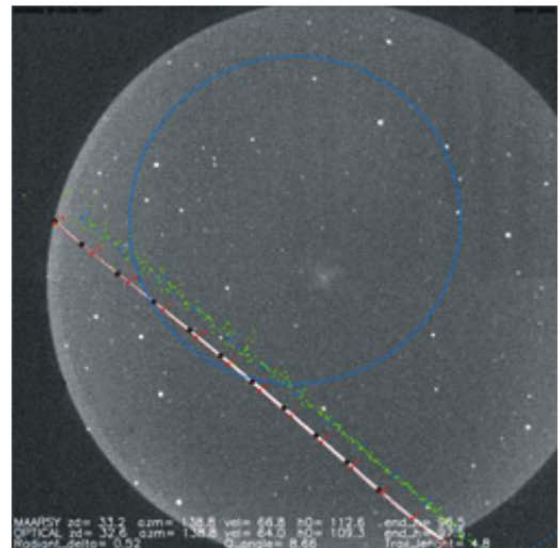

Figure 6. September 25, 2015 011302 UT event as seen by WATEC cameras (left) and narrow field intensified (left) cameras. Alomar is at the top and Saura at the bottom. The blue circle in each optical field represents the 3 dB points of the main MAARSY beam projected to 100 km altitude. Each green dot represents pulse-to-pulse interferometric tracking and projection on the plane of sky at the site using the actual head echo height measured from the MAARSY range and interferometry. The outer purple circle on the Watec images (left) represent the first null in the MAARSY radar gain beam.

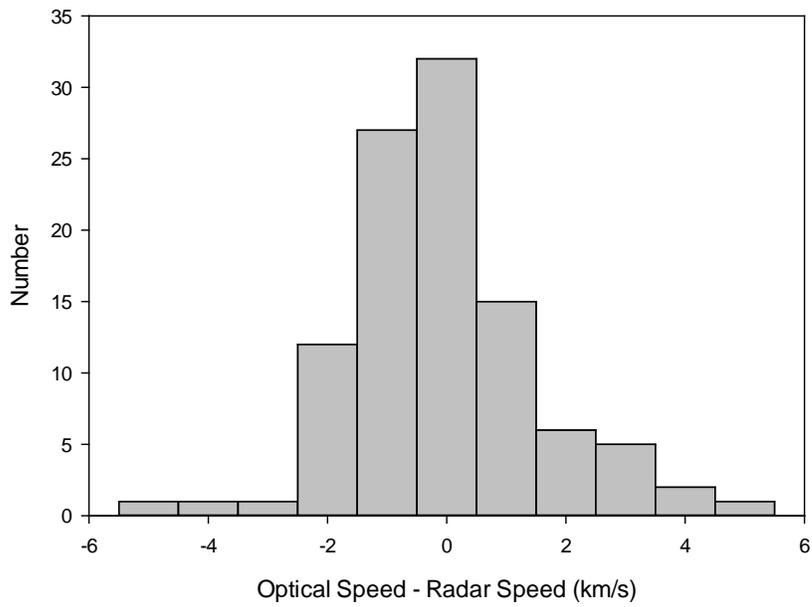

Figure 7. Average speed difference between common meteor events detected optically and as head echoes by MAARSY.

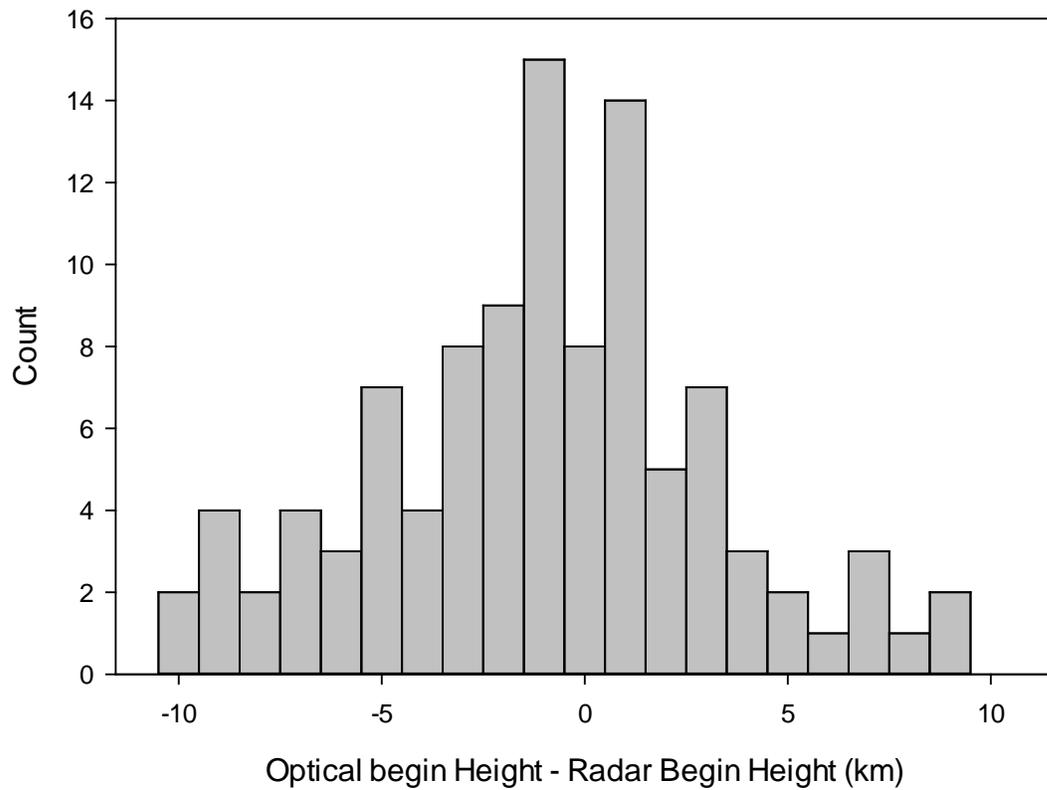
Figure 8. Average begin height difference between common meteor events detected optically and as head echoes by MAARSY.

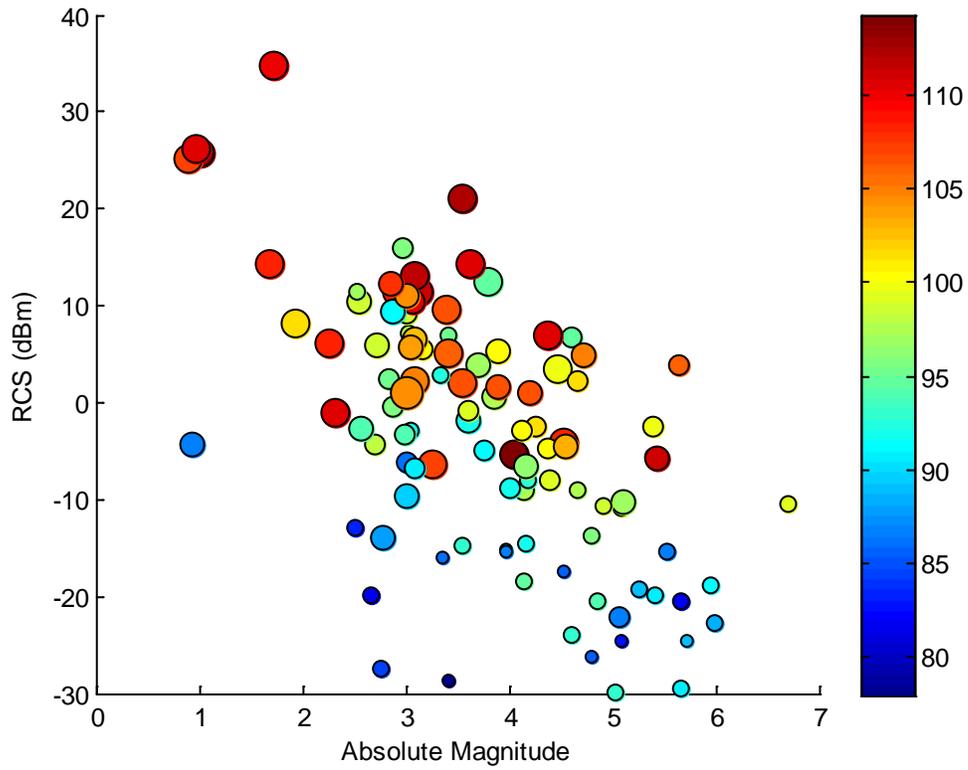

Figure 9. Peak radar cross section (in units of dB relative to a 1 m$^2$ target) versus the observed magnitude at the same height as the RCS measurement for the head echo with symbol size proportional to speed and color coding by height (in km).

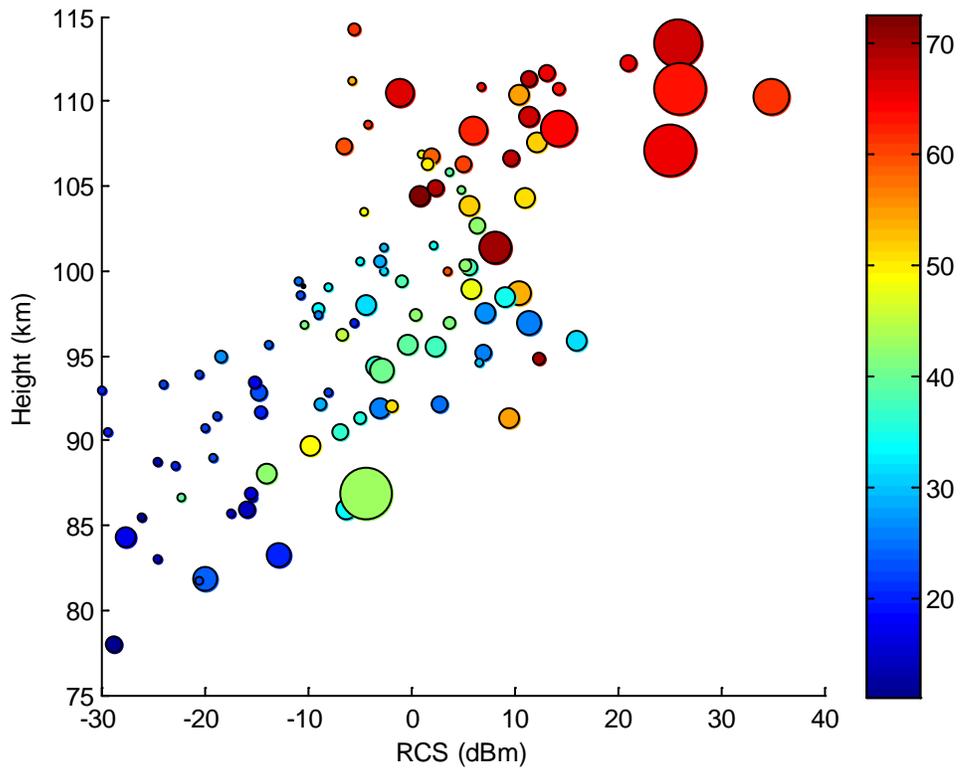

Figure 10. Height versus peak radar cross section (in units of dB relative to a 1 $m^2$ target) as a function of speed (color coding in km/s) with symbol sizes representing peak meteor absolute brightness in watts.

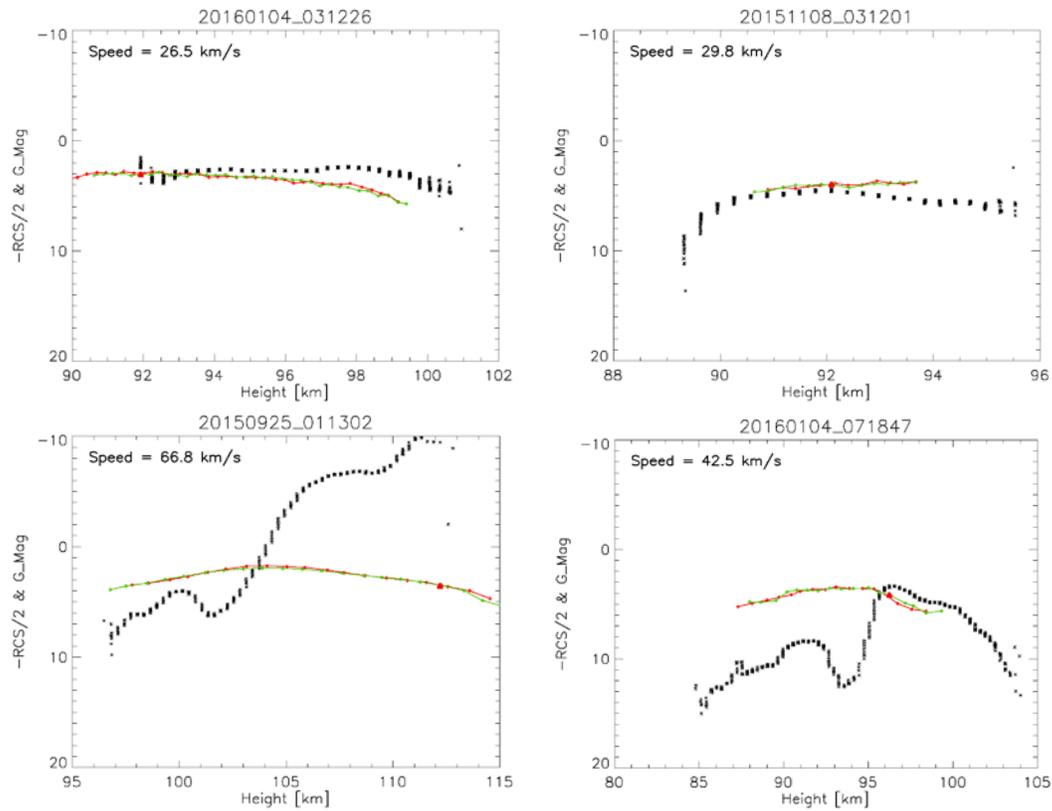

Figure 11. Radar cross section (in units of dB relative to a 1 m$^2$ target scaled as –RCS/2 on the y-axis so that larger RCS values are upward) shown as solid black dots versus height for four events. Also shown are the wide field (WATEC) absolute lightcurves in units of G-band magnitude (y-axis) from the Saura camera (red line) and Alomar (green line). The height on the optical lightcurve where the maximum RCS is reached is shown with a red triangle. The range of y-values is the same for all four events (ranging from 20 dBsm at the top to -40 dBsm at the bottom of the plot). The date and time of the event are shown in each title and the inset gives the meteor speed.

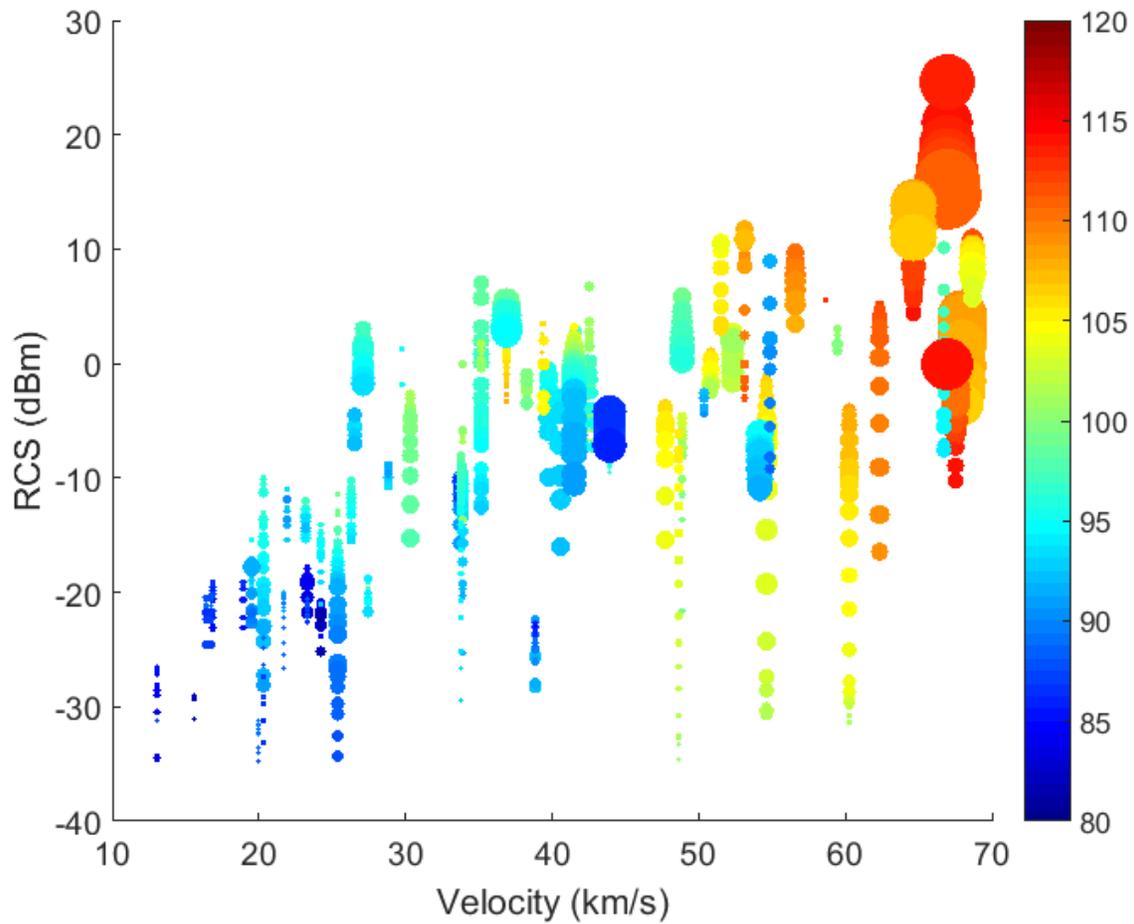

Figure 12. The RCS for 59 head echoes which fall within the main MAARSY beam (no pointe more than five degrees off vertical). Each point represent a single video frame where height, speed, rcs and magnitude are measured. Here the bright in watts is shown proportional to the size of the symbol and height (km) is colorcoded.

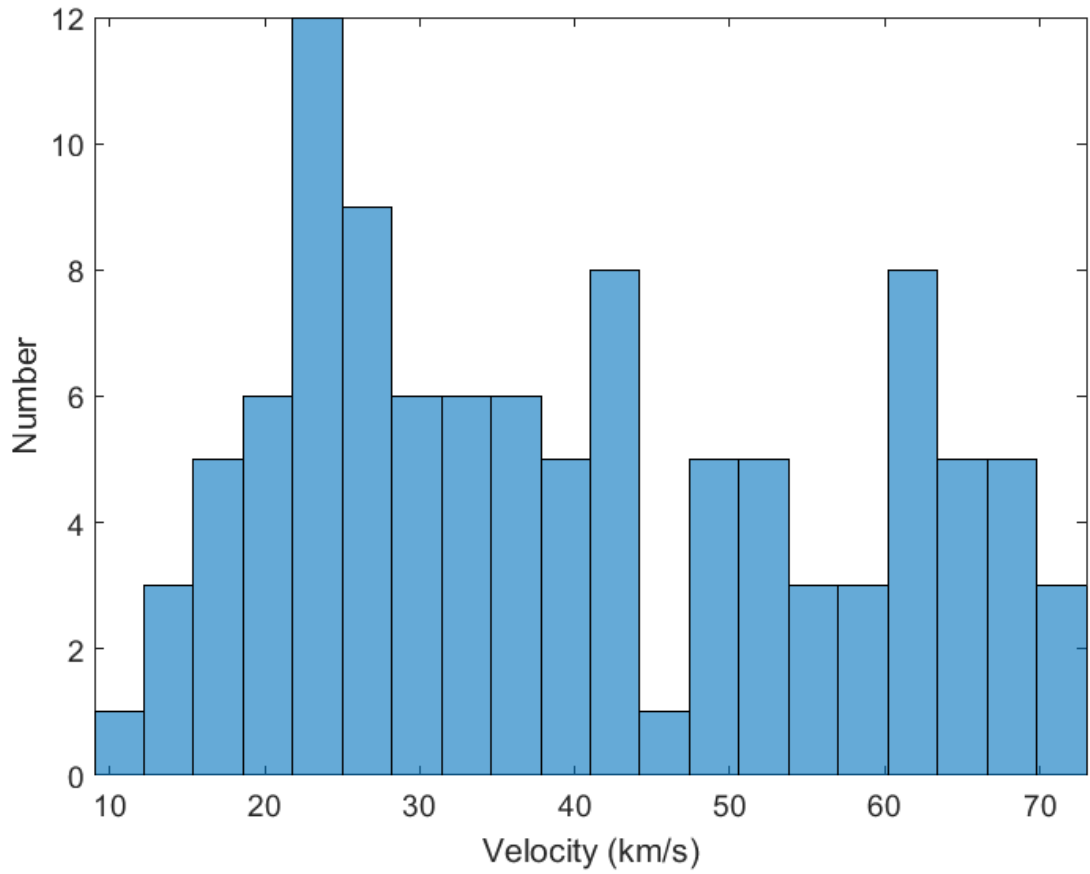

Figure 13. Velocity distribution of all 105 head echoes simultaneously detected optically.